
\input harvmac
\ifx\epsfbox\UnDeFiNeD\message{(NO epsf.tex, FIGURES WILL BE IGNORED)}
\def\figin#1{\vskip2in}
\else\message{(FIGURES WILL BE INCLUDED)}\def\figin#1{#1}\fi
\def\ifig#1#2#3{\xdef#1{fig.~\the\figno}
\goodbreak\midinsert\figin{\centerline{#3}}%
\bigskip\centerline{\vbox{\baselineskip12pt
\advance\hsize by -1truein\noindent\footnotefont{\bf Fig.~\the\figno:} #2}}
\bigskip\endinsert\global\advance\figno by1}

\overfullrule=0pt
\def\Title#1#2{\rightline{#1}\ifx\answ\bigans\nopagenumbers\pageno0\vskip1in
\else\pageno1\vskip.8in\fi \centerline{\titlefont #2}\vskip .5in}

\font\ticp=cmcsc10

\font\secfont=cmcsc10

%
%
\baselineskip=18pt plus 2pt minus 2pt

\def\ajou#1&#2(#3){\ \sl#1\bf#2\rm(19#3)}
%
\def\CH{{\cal H}}

\def\CF{{\cal F}}
\def\CS{{\cal S}}

\def\tN{{\tilde N}}

%


\def\z{ z}
%


\def\r{\rho}

\def\g{\gamma}
\def\t{\tau}
\def\a{\alpha}
\def\b{\beta}

\def\p{\pi}
\def\e{\epsilon}
\def\w{\omega}

\def\G{\Gamma}

%


\def\TrH#1{ {\raise -.5em
                      \hbox{$\buildrel {\textstyle  {\rm Tr } }\over
{\scriptscriptstyle \CH _ {#1}}$}~}}

\def\IZ{\relax\ifmmode\mathchoice
{\hbox{\cmss Z\kern-.4em Z}}{\hbox{\cmss Z\kern-.4em Z}}
{\lower.9pt\hbox{\cmsss Z\kern-.4em Z}}
{\lower1.2pt\hbox{\cmsss Z\kern-.4em Z}}\else{\cmss Z\kern-.4em Z}\fi}
\def\IC{\relax\hbox{$\inbar\kern-.3em{\rm C}$}}
\def\IR{\relax{\rm I\kern-.18em R}}
\def\1{\relax 1 { \rm \kern-.35em I}}
\font\cmss=cmss10 \font\cmsss=cmss10 at 7pt

\def\extraspace{\nobreak \hskip 0pt plus .15em\relax}
\def\dash{\unskip\extraspace--\extraspace}
%

\def\frac#1#2{{#1 \over #2}}
\def\ie{{\it i.e.}}

\def\p+{{\partial_+}}

\def\half{{1 \over 2}}

\def\imt{{\rm Im}\tau}
\def\ret{{\rm Re}\tau}

%


%

\Title{\vbox{\baselineskip12pt
\hbox{CALT-68-1953}
\hbox{\ticp doe research and}
\hbox{\ticp development report}
\hbox{}
\hbox{hep-th/9409158}
}}
{\vbox{\centerline {\bf QUANTUM CORRECTIONS TO BLACK HOLE ENTROPY}
  \vskip2pt\centerline{{\bf IN STRING THEORY} } }}

\centerline{{\ticp Atish Dabholkar}}
\vskip.1in
\centerline{\it Lauritsen Laboratory of  High Energy Physics}
\centerline{\it California Institute of Technology}
\centerline{\it Padadena, CA 91125, USA}
\centerline{ e-mail: atish@theory.caltech.edu}
\vskip .1in

\bigskip
\centerline{ABSTRACT}
\medskip

The one-loop contribution to the entropy of a black hole
from field modes near the horizon is computed in string
theory. It is modular invariant and ultraviolet finite.
There is an infrared divergence that signifies an
instability near the event horizon of the black hole.
It is due to the exponential growth of the density of
states and the associated Hagedorn transition characteristic
of string theory. It is argued that this divergence is
indicative of a tree level contribution, and the
Bekenstein-Hawking-Gibbons formula for the entropy should be
understood in terms of string states stuck near the horizon.

\bigskip

\bigskip
\Date{9/94}

\vfill\eject

\def\np#1#2#3{{\sl Nucl. Phys.} {\bf B#1} (#2) #3}
\def\pl#1#2#3{{\sl Phys. Lett.} {\bf #1B} (#2) #3}
\def\prl#1#2#3{{\sl Phys. Rev. Lett. }{\bf #1} (#2) #3}
\def\prd#1#2#3{{\sl Phys. Rev. }{\bf D#1} (#2) #3}

\def\cmp#1#2#3{{\sl Comm. Math. Phys. }{\bf #1} (#2) #3}

\lref\thooft{G. 't Hooft, \np{256}{1985}{727}.}
\lref\callwilc{C. G. Callan and F. Wilczek, \pl{333}{1994}{55},
hep-th/9401072.}
\lref\sussuglu{L. Susskind and J. Uglum,
{\it Black Hole Entropy in Canonical Quantum Gravity and \hfil\break
Superstring Theory}, Stanford Preprint SU-ITP-94-1 (1994), hep-th/9401070.}
\lref\bombetal{L. Bombelli, R. Koul, J. Lee and R. Sorkin,
\prd{34}{1986}{373}.}
\lref\srednick{M. Srednicki, \prl{71}{1993}{666}.}
\lref\kabastra{D. Kabat and M. J. Strassler, \pl{329}{1994}{46},
hep-th/9401125.}
\lref\bekenste{J. D. Bekenstein, \prd{7}{1973}{2333}; \prd{9}{1974}{3292}.}
\lref\hawking{S. W. Hawking, \prd{14}{1976}{2460};
\cmp{43}{1975}{199}; \cmp{87}{1982}{395}.}
\lref\gibbhawk{G. W. Gibbons and S. W. Hawking,
\prd{15}{1977}{2752}\semi
S. W. Hawking, \prd{18}{1978}{1747}.}
\lref\birrdavi{N. D. Birrell and P. C. Davies, {\it Quantum Fields in
Curved Space}, Cambridge University Press (1982).}
\lref\cone{A. Dabholkar,
{\it String Theory on a  Cone and Black Hole Entropy},
Harvard University Preprint HUTP-94-A019 (1994), hep-th/9408098}
\lref\dowker{J. S. Dowker,  {\sl Class. Quant. Gravity} {\bf 11} (1994) L55,
hep-th/9401159.}
\lref\teitelbo{M. Ba{$\tilde {\rm n}$}ados, C. Teitelboim  and J. Zanelli,
\prl{72}{1994}{957}, hep-th/9309026.}
\lref\carlteit{S. Carlip, C. Teitelboim, {\it The off Shell Black Hole}, IAS
preprint
IASSNS-HEP-93-84 (1993), gr-qc/9312002.}
\lref\susskind{L. Susskind, {\it Some Speculations about Black Hole Entropy
in String Theory}, Rutgers University preprint RU-93-44 (1993),
hep-th/9309145.}
\lref\fioletal{T. M. Fiola, J. Preskill, A. Strominger, S. P. Trivedi,
{\sl Black Hole Thermodynamics and Information loss in Two Dimensions},
Caltech Report CALT-68-1918 (1994), hep-th/9403137.}
\lref\alvaoso{E. Alvarez  and M. A. R. Osorio, \prd{36}{1987}{1175}.}
\lref\polchinski{J. Polchinski, \cmp{104}{1986}{37}.}
\lref\macroth{B. Maclain and B. D. B. Roth, \cmp{111}{1987}{539}.}
\lref\obritan{K. H. O'brian and C. I. Tan, \prd{36}{1987}{477}.}
\lref\aticwitt{J. Atick and E. Witten, \np{310}{1988}{291}.}
\lref\rohm{R. Rohm, \np{237}{1984}{553}.}
\lref\hagedorn{R. Hagedorn, Nuovo Cimento Suppl. {\bf 3} (1965) 147\semi
E. Alvarez, \prd{31}{1985}{418}\semi
B. Sundborg, \np{254}{1985}{583}\semi
S. H. Tye, \pl{158}{1985}{388}\semi
M. J. Bowick and L. C. R. Wijewardhana,  \prl{54}{1985}{2485}.}
\lref\satkog{B. Sathiapalan, \prd{35}{1987}{3277}\semi
Ya. I. Kogan {\sl JETP Lett.} {\bf 45} (1987) 709.}
\lref\inprep{{\sl in preparation}.}
\lref\greenbook{M. B. Green, J. H. Schwarz and E. Witten,
{\it Superstring Theory},  {\rm vols. I and II},
Cambridge University Press (1987).}
\lref\membrane{K. S. Thorne, R. H. Price, and D. A. MacDonald,
{\it Black Holes: The Membrane Paradigm}, Yale University Press (1986),
and references therein.}
\lref\stretched{L. Susskind, L. Thorlacius and J. Uglum,
\prd{48}{1994}{3743}, \hfil\break hep-th/9306069.}
\lref\thornetal{K. S. Thorne and D. A. MacDonald,
{\sl Mon. Not. Roy. Astron. Soc.} {\bf 198} (1982) 339\semi
W. H. Zurek and K. S. Thorne, \prl{54}{1985}{2171}\semi
R. H. Price and K. S. Thorne, \prd{33}{1986}{915}.}
\lref\membone{R. S. Hanni and R. Ruffini, \prd{8}{1973}{3529}\semi
P. Hajicek, \cmp{36}{1974}{305}\semi
R. L. Znajek, {\sl Mon. Not. Roy. Astron. Soc.} {\bf 185} (1978) 833.}
\lref\barbon{J.~l.~F.~Barb\'on, {\it Remarks on Thermal
Strings Outside Black Holes}, Princeton University Preprint
PUPT-94-1478 (1994), hep-th/9406209.}
\lref\bowgid{M. J. Bowick and S. B. Giddings, \np{325}{1989}{631}.}
\lref\devsan{H.J de Vega and N. Sanchez, \np{299}{1988}{818};
\prd{42}{1990}{3969}; N. Sanchez, \pl{B195}{1987}{160}.}
\lref\susstwo{L. Susskind, \prd{49}{1994}{6606};
\prl{71}{1993}{2367}.}

\newsec{Introduction}

A major  puzzle in the physics of black holes concerns the  interpretation
of the entropy associated with a black hole.
In the semi-classical approximation, the entropy is given by the
Bekenstein-Hawking formula \refs{\bekenste , \hawking}:
\eqn\Sbh{S_{BH}  = {A \over {4 G \hbar}} \ ,}
where A is the area of the event horizon of the black hole and
$G$ is Newton's constant.
This remarkable formula points to a  deep connection between quantum
mechanics, gravity and  thermodynamics.
However, its relation to statistical mechanics is largely mysterious.
Gibbons  and Hawking \gibbhawk\  have obtained
the same expression from the quantum partition function for gravity.
Even so, a  satisfactory statistical interpretation of the entropy
in terms of enumeration of states is lacking.

Another related puzzle concerns  the entropy of
quantum fluctuations seen by the Schwarzschild
observer. As pointed out by 't~Hooft \thooft ,  the contribution
to the entropy coming
from the field modes very near the horizon is ultraviolet divergent.
Several authors  have found a similar divergence in Rindler spacetime
which approximates the geometry of a large black hole very near the horizon
\refs{\sussuglu ,\srednick, \dowker , \bombetal , \callwilc , \kabastra}.
There is a simple description of the leading divergence.
A fiducial observer that is stationed at a fixed radial distance from the
black hole, has to accelerate with respect to the freely
falling observer in order not to fall into the black hole.
Very near the horizon, the fiducial observer is like a Rindler observer
in flat Minkowski space.
As a result, she sees a thermal bath \birrdavi\ at a
position-dependent proper temperature  ${ T(\z )} = {1\over{2\pi \z}}$ where
$z$ is the proper distance from the horizon.
Using Planck's formula for a single massless boson
we get the entropy density:
\eqn\fourentropy{
s(z) = {4\over 3} {\pi^2 \over 30} ({1\over{2\pi \z}})^3\,  .}
Note that we have been able to define the entropy density because
entropy is an {\it extensive} quantity as it should be.
However,  the dominant contribution comes from the region
near the horizon $z=0$ and is not extensive but proportional
to the area.   If  we put a cutoff on the proper distance  at
$z=\e$ (or alternatively on proper temperature) the total entropy is:
\eqn\totalentropy{\eqalign{
S &= \int _\e^\infty { s(z) A dz}\cr
   &=  {A \over {360\pi \e^2}}  \quad ,\cr}
}
where $A$ is the area in the transverse dimensions.
This is in agreement with the result obtained in
\refs{\thooft , \sussuglu}\  by other means.
Because the thermal bath is obtained by tracing over states that
are not accessible to the observer in the  Rindler wedge,
this entropy is  also the same as the `entropy of
entanglement'  \refs{\bombetal , \srednick , \kabastra ,  \fioletal }
or the `geometric entropy'  \refs{\dowker , \callwilc }.
For a massive particle of mass $m$ there will be corrections to
this formula which will be down by powers of $m\e$.

't Hooft has advocated that the entropy of entanglement should account
for the black hole entropy. This would offer a statistical interpretation
of the black hole entropy and also a natural explanation for
the area dependence.
In the context of field theory, there are several difficulties
with this appealing idea.
For example, the entropy of entanglement is divergent and
depends on an arbitrary  ultraviolet cutoff
whereas the black hole entropy is finite and depends on the
Planck length. The entropy of entanglement depends on the species
and the couplings of the various particles in the theory
whereas the black hole entropy is
species-independent. Finally, it is difficult to see how
the entropy of entanglement, which always starts at one loop can
possibly account for the black hole entropy which is inversely
proportional to the coupling constant.
't Hooft has argued that it is necessary to understand the
ultraviolet structure of the theory in order to address these issues.
He has conjectured that these difficulties will be resolved
once the correct short distance structure is known.
He has further suggested that this divergence of entropy in
field theory is intimately
related to the puzzle of loss of information in black hole evaporation.
If the entropy does have a statistical interpretation in terms of
counting of states, then its divergence would suggest an infinite
number of states associated with  a finite mass black hole.
As long as the black hole has an event horizon, it can apparently store
an arbitrary amount of information in terms of correlations
between the outgoing radiation and the high energy modes near
the horizon. When the horizon eventually disappears,
the information in these correlations is irretrievably lost.

This conjecture could hardly be tested in field theory especially
when we are dealing with a non-renormalizable theory such as
quantum gravity.
Fortunately, string theory offers a suitable
framework for addressing this question.
It is a perturbatively finite theory of quantum gravity
and comes with a well-defined matter content.
Moreover, Susskind \refs{\susstwo, \susskind}
has argued that string theory
may also possess some of the properties required for describing
black hole evaporation without information loss.
It is therefore of great interest to know how
the ultraviolet behavior of the entropy is controlled
in string theory.
Until now, the finiteness of string theory
has been little utilized in quantum gravity.
The main obstacle has been that
many of the difficult questions in quantum gravity require
a  non-perturbative formalism of the theory.
By contrast, the question about the
ultraviolet finiteness of entropy  is accessible entirely
within perturbation theory.
One can hope that string theory
will illuminate this question in important ways.

In the next section,  we compute the one-loop contribution
to the entropy of a very large black hole
in the twenty-six dimensional bosonic string.
We obtain a modular invariant expression which easily
generalizes to higher loops. The  generalization
to superstrings is also straightforward.
The main result of  this paper is that
the quantum contributions to the entropy of a very large black hole in
string theory  are  indeed finite in the ultraviolet.
This finiteness comes about by a mechanism well-known
in string theory. In field theory, the divergence comes from summing over
small loops  in which the virtual particles travel for a very short
proper time.
In string theory, this region of short proper time belongs to certain
corners of the parameter space that is excluded by modular invariance.

It turns out that the entropy,  even though finite in the ultraviolet,
is divergent in the infrared.
The divergence is  due to a  tachyonic instability
very near the horizon and most likely is associated with the Hagedorn
phase transition \hagedorn .
Recall that the fiducial observer sees a  hot thermal bath
near the horizon. Moreover,  the density
of states at high energy has an exponential
growth characteristic of string theory.
As a result, beyond a certain temperature \dash the so-called Hagedorn
temperature \dash the thermal ensemble is rendered unstable.
If there is a phase transition and
a condensate is formed, then the latent heat
of the transition can induce a tree level contribution
to the entropy \aticwitt . Thus, it seems plausible
that we can understand the dependence of the entropy on coupling
constant. Moreover, if the contribution to the entropy comes
from the fundamental degrees of freedom of string theory, then
quite possibly it is independent of the species of particles
in the low energy theory.
Finally, as we shall see, in the case of black holes,
this condensate will be restricted
very close to the horizon and will extend
only in the transverse directions.
There can also be an additional contribution to the entropy
from the string states  with endpoints immersed in the
`deconfined' soup of strings very near the horizon \susskind .
This raises the exciting possibility that the  area dependence of the
tree level  Bekenstein-Hawking
formula for the entropy can be understood in terms of statistical
mechanics.

\newsec{Black Hole Entropy in the Bosonic String at One Loop}

In this section we derive the expression for the entropy in
the bosonic string theory at one loop.
This derivation is essentially an application of equations
\fourentropy\ and \totalentropy .
These considerations are also relevant to Rindler strings.
Some aspects of Rindler strings have been discussed earlier
in \devsan .

As our starting point we take the expression for the entropy in field theory
in the proper time formalism.  The reasoning  leading to the corresponding
expression in  string theory is similar to the one employed in
the derivation of the cosmological constant \rohm .
Let us consider the  free-energy  density
for a single boson  of mass $m$ at finite temperature $\b^{-1}$:
\eqn\freeone{
f(\b ,m^2 ) = {1\over \b} \int \frac{d^{d-1} k}{(2 \pi )^{d-1} }
\log (1-e^{- \b \w_k} ).
}
To write it in the proper time formalism we first introduce
$1= \int_0^\infty 2 \w \int_{-i \infty}^{i \infty} \frac{ds}{2\pi i}
\ e^{s (\w^2 -\w_k^2 )}$,
expand the logarithm and then  perform the gaussian momentum integrals to
obtain \refs{\polchinski ,\macroth }
\eqn\freetwo{
f(\b , m^2 ) = -\int_0^\infty \frac{ds}{s} \frac{1}{(2\pi s )^{d/2}}
\sum_{r=1}^{\infty} e^{-m^2 s/2 - r^2\b^2/{2s} }.
}
The entropy density is as usual
$ s(\b , m^2 ) = \b^2 \frac{\partial f}{\partial \b}$.
For Rindler observers  we simply
put  $\b = 2 \pi z$ to obtain the local entropy density
and then integrate as in \totalentropy\ to get the total entropy:
\eqn\entropyone{
S(m^2 ) = A\int_{0}^{\infty} dz \left( 2\pi z \right)^3
\int_0^\infty \frac{ds} {s^2}\frac{1}{(2\pi s )^{d/2}}
\sum_{r=1}^{\infty} r^2 e^{-m^2 s/2 - 2\pi^2 r^2 z^2/{s} }.
}
There is an  ultraviolet divergence as  in \totalentropy\ because the
$ s$ integral diverges near $s=0$ for small $ z$.
Notice that we have to be careful while interchanging the order of
integration because the integral over $ s$ is not
uniformly convergent as a function of $ z$.
We can interchange the order by putting
appropriate cutoffs for both the integrals.

It is straightforward to generalize these formulae to  the spectrum of
the bosonic string  in $d= 26$ by summing the expression \freetwo\ over $m^2$.
In the light cone gauge the spectrum is given in terms of the
occupation numbers of the right-moving and the left-moving
oscillators $N_{ni}$ and $\tN_{ni}$ \greenbook :
\eqn\spectrum{
m^2 =\frac{2}{\a '} \, \left[ -2 + \sum_{i=1}^{24}\sum_{n=1}^{\infty} n
\left( N_{ni} + \tN_{ni} \right) \right],
}
subject to the constraint
\eqn\constraint{
\sum_{i=1}^{24}\sum_{n=1}^{\infty} n \left( N_{ni}
-  \tN_{ni} \right)  = 0.}
The constraint can be enforced by introducing
\eqn\con{
\int_{-\half}^{\half} d\theta\,
e^ {2\pi i n \theta  ( N_{ni} -  \tN_{ni} ) }
}
into the sum. It is convenient to
introduce a complex variable $ \tau = \theta  + i \frac{s}{2\pi \a '}$.
The sum can then be easily performed to obtain
\eqn\freethree{
f(\b ) = -\half (\frac{1}{4\pi^2 \a '})^{-13}\int_{\CS} \frac{d^2 \tau }
{  \imt ^{2}}{\left( \imt \right)^{-12}}  | \eta (e^{2\pi i \t} ) |^{-48}
\sum_{r=1}^{\infty} exp { -\frac{r^2\b^2}{4\pi \a ' \imt} }.
}
Here $ \eta $ is the Dedekind eta function,
\eqn\dedekind{
 \eta (q) = q^{\frac{1}{24}} \prod_{n=1}^{\infty}{(1-q^n)}
}
and  the region of integration is  the strip $\CS$
\eqn\strip{
-\half \leq \ret \leq \half , \quad 0 \leq\imt \leq \infty .}
 The total entropy computed from this formula has a divergence for
each  mode coming from the region near $\imt = 0 $.

This would be the end of the story if we were dealing  with a field theory
of the string modes.
But string theory is
not merely a sum of field theories because of duality \polchinski .
The sum of field theories overcounts the correct string answer.
For example, a  four point amplitude
is represented by  a single string diagram but
gives rise to two separate diagrams in the $ s$ and the $ t$ channel
in field theory.  The correct generalization of the above formulae
to string theory is more subtle and requires a proper treatment of this
overcounting. We  do this by noting that using a modular transformation of
$ \t$, every point in  $ \CS$ can be mapped
onto the fundamental domain $ \CF$ of a torus \macroth :
\eqn\funda{
|\t|> 1 \, ,\,  -\half < {\rm Re} \t < \half  \, , \, {\rm Im} \t > 0 \, .
}
Recall that the modular group $ \G$ at one-loop  is the group
of disconnected diffeomorphisms of  a torus up to conformal
equivalences. It is isomorphic to the group $  SL(2, \IZ) / Z_2$
under which $ \t$ transforms as
\eqn\modtwo{\t '= \frac{a\t +b}{c\t +d} \,  , \qquad \quad
\pmatrix{
a & b \cr
c & d \cr
} \epsilon\ SL(2, \IZ) .
}
We have to divide the $SL(2,  \IZ )$ by $Z_2$ because the elements
$\{ \1 , -\1\} $ leave
$\t$ unchanged.
The strip $ \CS$ consists of an infinite number of domains $ \CF_\g$
each of which can be obtained from $ \CF$ by the action
of an element  of $ \G$,
$ \g (\r_1 \r_2 ) = \pmatrix{
a & b \cr
\r_1 & \r_2 \cr
}$ where $ \r_1 $ and $ \r_2$ are relatively prime.
The integers $ a $ and $ b$ can be chosen in such
a way that $ \imt '  = \imt  /E\left( \t, \r_1 , \r_2 \right)$
where  we have defined
\eqn\noname{
E\left( \t, \r_1 , \r_2 \right) \equiv | \r_1 \t  + \r_2 |^2 .}
Note that in expression \freethree ,  if we replace the summation by $ 1$
then what we have is the  cosmological constant
at one loop for the bosonic string at zero temperature \greenbook\ which
is invariant under modular transformations.
Using these facts we see that
\eqn\equal{\eqalign{
f(\b ) &= -\half (\frac{1}{4\pi^2 \a '})^{-13}\int_{\CS} \frac{d^2 \tau ' }
{ { \imt  '}^{2}}
{\left( \imt ' \right)^{-12}}  | \eta (e^{2\pi i \t '} ) |^{-48}
\sum_{r=1}^{\infty} exp { -\frac{\b^2r^2}{4\pi \a ' \imt '} } \cr
 &= -\half (\frac{1}{4\pi^2 \a '})^{-13}\int_{\CF} \frac{d^2 \tau }
{  \imt ^{2}}
{\left( \imt \right)^{-12}}  | \eta (e^{2\pi i \t} ) |^{-48}
\sum_{r_1, r_2}
exp { -\frac{\b^2 E\left( \t, r_1 , r_2 \right)}{4\pi \a ' \imt} }. \cr
}}
It follows now that the total entropy in the bosonic string at one loop is
given by
\eqn\entropy{\eqalign{
S = (\frac{1}{4\pi^2 \a '})^{-13}
2\pi A & \int_{0}^{\infty} dz\left( 2\pi z \right)^3
 E\left( \t, r_1 , r_2 \right)  \times\cr
& \times\int_{\CF} \frac{d^2 \tau }{  \imt ^{2}}
{\left( \imt \right)^{-13}}  | \eta (e^{2\pi i \t} ) |^{-48}
\sum_{r_1, r_2}
exp { -\frac{\pi z^2 E\left( \t, r_1 , r_2 \right)}{ \a ' \imt} }.\cr
}}
It is easy to check that this expression is modular invariant using
the Poisson resummation formula. This means that the restriction of
the modular integration to the fundamental domain $\CF$ is a
consistent procedure.
\ifig\fone{In string theory the modular integration is over the
fundamental domain $\CF$ and not over the strip $\CS$. The entropy
is ultraviolet finite because the region of short proper time near
$\imt =0$ is excluded by modular invariance.
Infrared divergences coming from very large $\imt$ may still be present.}
{\epsfysize=2.6in \epsfbox{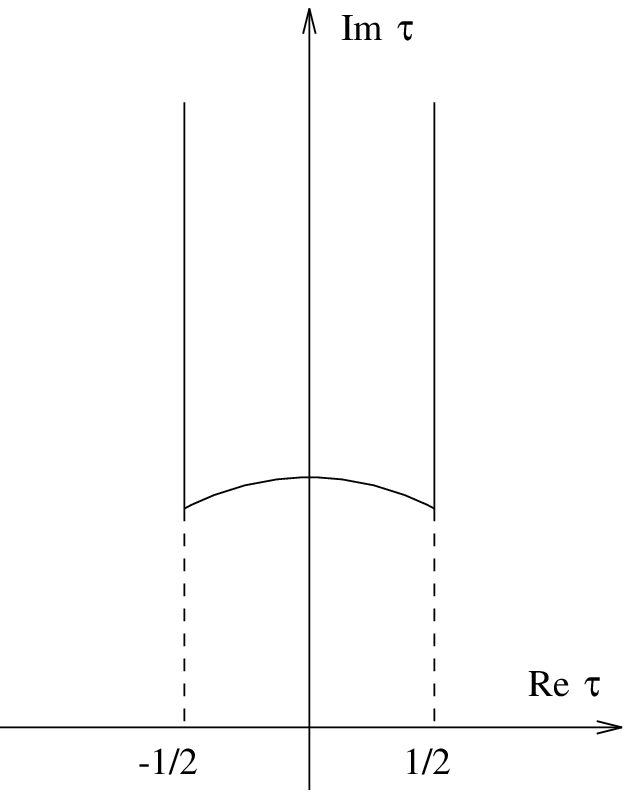}}
\noindent The most important consequence
of modular invariance is that the entropy is ultraviolet finite.
The troublesome region in the proper time integration
that gave rise to the divergence in field theory is
excluded in string theory (see \fone ).
By performing  the $z$ integral first, we can obtain the entropy
as an integral of a density over the fundamental domain.

A striking property of the above formula is that it suffers from a
severe infrared divergence. The integral diverges when
$ z$ is small and $\imt$ is large which corresponds
to virtual particles spending a long proper time
near the horizon.
The interpretation of this divergence is not completely clear
to us. One possibility is to render this expression finite by appropriate
deformation of the $ z$ contour. Note that the expression is modular invariant
for an arbitrary contour in the complex $ z$ plane.
By shifting the contour away from the real axis it is possible
to  get a finite expression.
This is analogous to the situation in statistical mechanics when
a partition function has a pole for real $\b $.
One still obtains an exponentially increasing but finite density of states
because the proper choice of contour of the inverse Laplace transfom
avoids the singularity.
Another possibility is that the $ z$ integral is defined via a principal
value prescription.
While both these prescriptions would lead to  mathematically well-defined
expressions, their physical justification is far from clear.
{}From a physical point of view,
it seems much more plausible that the divergence is indicative
of a Hagedorn-like  transition
\refs{\hagedorn , \satkog , \obritan , \aticwitt }\
well-known in string theory.
We can gain some insight by rewriting  the formula \entropy\ in
a more suggestive form. Using the Poisson resummation formula,
it is easy to see that
\eqn\radius{
\sum_{r_1, r_2}
exp { -\frac{\pi z^2 E\left( \t, r_1 , r_2 \right)}{ \a ' \imt} }
=\frac{1}{2\pi z} \sqrt{4\pi \a ' \imt }\sum_{m, n}
e^{-2\pi m n \ret} e^{-\pi \imt
\left( {\a ' m^2}/ {z^2} +{n^2 z^2}/{\a '}  \right)}.
}
With this equality we make contact with the work of Sathiapalan and Kogan
\satkog . They observed that the Hagedorn transition can be
identified with the appearance of a tachyon that comes
from  a winding mode in the compactified Euclidean time direction.
The radius of the circle in the time direction is $ R = z /{\sqrt{\a '}}$.
In the above formula,  the integer $ n$ can be identified with the
winding number. To look for the tachyonic divergence we look for terms
that go as $ e^{ + \imt }$ at large $\imt$ which then produce a divergence
from the modular integration.
For example, if we consider the term with $ m=0$ and $ n =\pm 1$ we find  that
there is no tachyonic divergence for large $z$ but it first appears at
$z= \sqrt{2\a '}$. This is only the first tachyonic mode
that appears as we lower $z$.
At  smaller and smaller values of $ z$ (\ie ,
at higher and higher temperatures),
more and more winding modes become tachyonic.
The exponential growth of the density of states gives rise
to poles at these various values of $z$.
This means that we must really abandon the thermal ensemble for
this region of the $ z $ integral.
It is worth pointing out in this
context  that in theories that have the $R\rightarrow \frac {1}{R}$ duality,
like the heterotic string, there is no divergence
at $ z=0$ but only at intermediate values of $ z$.

\newsec{Discussion}

It is equally easy to compute the
one loop entropy in superstring theory, by using the
known formulae for superstrings at finite temperature
\refs{\obritan ,\alvaoso ,\aticwitt}.
Once again we obtain a modular invariant
expression for the entropy that is ultraviolet finite.
The infrared divergences are still present, and they should be.
Superstrings do not have a tachyon in the spectrum  to begin with.
Nevertheless, they have an exponentially growing density of states at
high energy and the corresponding Hagedorn transition.
The tachyons that are relevant here  have to do with this
phase transition and they persist even in the case of superstrings.

It may appear that we have simply traded an ultraviolet divergence for an
infrared divergence without gaining anything.
After all, we started with the expression in field theory and rewrote it.
However, it should be remembered that
our ability to restrict the region of integration to the fundamental domain
depended crucially on the very specific spectrum of string theory.
It would not be possible for an arbitrary collection of field
theories.
Moreover, the physical interpretation
of these two divergences is profoundly different.
An ultraviolet divergence is usually indicative of new degrees of
freedom or a plain inconsistency of the theory.
For example, the ultraviolet divergences in the Fermi theory
of weak interactions are indicative of the existence of massive
vector bosons. After adding these new degrees of freedom, the ultraviolet
divergences are removed from the more
complete theory. In this sense,
the ultraviolet divergences are unphysical and are a feature only of
the low energy theory.
Infrared divergences, on the other hand, contain important
physics. In the present context,
the infrared divergence is indicative of an instability and
possibly a phase transition.
It says that we have not treated the tree level
answer correctly.
Ordinarily, in string theory, the tree level free energy is zero
because the sphere partition function vanishes by conformal invariance.
Here, a tree level contribution
to the entropy will be induced by the
latent heat of the phase transition.

All our reasoning so far has been from a Hamiltonian point of view.
It would be most interesting to obtain the same results from a
Euclidean path intgral \gibbhawk .
The Euclidean continuation of Rindler spacetime is simply
flat space with the angular variable in a plane
playing the role of Euclidean time. The periodicity $2 \pi $
of the angular variable is proportional to the inverse temperature $\b$.
In order to compute the entropy we need to take a derivative
of the free energy with respect to $ \b$ . Consequently we need
to understand string propagation on a plane
where the periodicity  of the angular variable differs
from $2\pi $. This amounts to studying string
theory on a cone with some deficit angle
\refs{\susskind ,\sussuglu  ,\cone}.
As described in \cone ,
the twisted sectors play an important role
in the formulation of string theory on a conical orbifold.
It seems plausible that the twisted states and the
winding modes above are somehow related. Both
are necessary in string theory to ensure modular
invariance but are difficult to describe in the
corresponding field theory.

A path integral derivation is desirable for another reason.
In this paper, we derived the entropy by integrating a local density.
This is certainly correct at one loop. At higher loops, the local
entropy density may not be very well-defined
because the thermal wavelength at a given position
is of the same order as the proper distance from the horizon.
We wish to return to a more complete comparison of the two
approaches in a subsequent publication
\inprep .
We expect that both  these methods for computing  the entropy
should agree in the end, at least at one loop.
The necessity for including the winding modes means that the Euclidean path
integral for Rindler spacetime should be regarded as a path integral
not over a plane, but over a plane with the origin removed.
A particle does  not notice  the difference between the two because
there is an ultraviolet divergence near the origin which
has to be regulated.  For a string, there are
no ultraviolet divergences. However, the string knows that
it is  moving not on  a plane but on a topologically nontrivial space and
hence the winding modes must be included.
Notice that the variable $z$ above measures the radial distance
away from the origin.
Each winding mode at some radius $ z$ represents a state that
has a position dependent mass-squared which gets smaller towards the origin
and eventually becomes tachyonic.

At this stage, we would like to mention the various approximations
that we have used in this calculation:

1. We have worked in the limit of infinite black hole mass.
There will certainly be finite mass corrections to the black
hole entropy. At present, we do not know how to include
these into our calculation.

2. We have ignored finite mass corrections to the entropy in
equation \totalentropy . These are down by powers of $m\e$
where $m$ is the mass of the quantum field and are unimportant
as long as $\epsilon$ is strictly zero. Once we have
an infrared divergence, we will have to keep $\e$ small
but nonzero. In that case, we will have to include the
finite mass corrections at some mass level. It is possible to
circumvent this difficulty by using the full heat kernel for
massive particle on a cone directly \inprep . Moreover, for
special values of the Rindler temperature we can compare
our results with the conical orbifold \cone .
We expect that many of the qualitative features such as
the existence of the Hagedorn transition etc., will not
be altered.

3. There is a more conceptual question regarding the validity
of using the thermal ensemble. A Rindler observer
sees a hot thermal bath and the thermal ensemble is not well-defined
in string theory \aticwitt . Moreover, once we include interactions
we also have to worry about the Jeans instability \bowgid .
It is not clear how to properly take these effects into account.
However, we are really interested only in the entropy at a
special value of the temperature and
we hope that at least some of the features of the entropy will
be accessible  without having to understand all the consequences
of the Hagedorn transition. For example, it would be nice
to see if the entropy can be rendered finite in the infrared
by adding a tree level contribution.

Susskind and Uglum \sussuglu\ have argued
that renormalization of entropy  can be absorbed into
the renormalization of Newton's  constant.
Their arguments are based on the analysis of ultraviolet
divergences in the low energy field theory.
Now consider a theory with enough supersymmetries
so that there is no renormalization of Newton's
constant at one loop.
Then, according to Susskind and Uglum, the black hole entropy is
given by the Bekenstein-Hawking formula at tree level
and there are no loop corrections.
The existence of the tree-level answer itself is
not explained by this renormalization procedure.
In the picture presented in this paper, the entropy is
naively zero at tree level.
There are no ultraviolet divergences in the full string theory.
Nonetheless, there are infrared divergences and
the tree-level entropy will be induced by the Hagedorn transition.
In this case, the arguments of Susskind and
Uglum imply that the induced tree-level entropy must {\it exactly}
equal the Bekenstein-Hawking entropy.

If this  picture turns out to be correct, we may be able
to learn something about the Hagedorn transition from its
relation to the black hole entropy and vice versa.
The Bekenstein-Hawking  formula can thus be viewed as
a valuable link to the
fundamental degrees of freedom of string theory.
The black hole entropy will then have a statistical interpretation and
the condensate near the horizon can provide a concrete realization,
in the context of string theory,
of the phenomenological idea of a ``membrane'' or a ``stretched horizon''
\refs{\membrane ,\membone ,\thornetal ,\stretched}.

\eject
\bigskip
\leftline{ \secfont Acknowledgements}
\bigskip
I would like to thank C.~Vafa and F.~Wilczek for discussions,
and J.~Gauntlett, J.~Preskill, and L.~Susskind for comments
on the manuscript.
This work was supported in part by the U. S. Department of Energy
under Grant No. DE-FG03-92-ER40701.

Note added: After completing this work, I learned of a preprint
by J.~L.~F.~Barb\'on \barbon\ in which the thermal ensemble of free strings
outside black holes is discussed with the `brick wall'
horizon regularization of 't~Hooft.

\listrefs
\end